\documentclass[showpacs,prl]{revtex4}
\begin{document}
\title{ Coherent lunar effect on solar neutrino:double slitt interference } 
\author{ K. Ishikawa and  T.Shimomura  }
%\maketitle
\affiliation{  Department of Physics, Hokkaido University, 
             Sapporo 060-0810, Japan }

%  Department of Physics, Faculty of Science, Hokkaido University
%Sapporo 060-0810, Japan

%\date{\today}

\begin{abstract}
Coherent interaction of solar neutrino with the moon and its implication 
are investigated. We solve the Dirac equation with the moon potential
and show that a phase shift of the neutrino wave function becomes almost 
one unit if the neutrino penetrates the moon. Spatial interference is
 generated for the neutrino wave packets during eclipse and possibility of  
observing interference effect that is similar to that of a double slit 
experiment is pointed out.

\end{abstract}
\maketitle
%\section{Introduction}
%\subsubsection{Key  words}
%neutrino astronomy, study  of planet by solar neutrino,
%incoherent neutrino interaction and a mean  free path of neutrino in the 
%earth,   coherent interaction, double slit neutrino experiments, reflaction 
 
Since  neutrino from Supernova were observed [1], neutrino astronomy
started. An astronomical observation using neutrino could supply a new 
information which is not  obtained by  light. However neutrino interacts 
with matter so weakly that event rate is extremely small. High
luminosity and enhancement of event rate are required to make a real 
observation possible.
The most luminous source of neutrino around the earth is sun and it is
practical to use solar neutrino for astronomical purpose. Especially
solar neutrino of discrete energy from $Be^7$ process can be a useful 
source for astronomical observations. Coherent scattering of neutrino by 
neutral current interactions can enhance the event rate.  We 
study a coherent moon effect to solar neutrino in this paper. 

Coherent interaction of neutrino with  massive objects is a unique macroscopic
quantum phenenomenon which appears despite an extreme weakness of interaction
with matters. The flux of solar neutrino is $10^{9} /{cm^2sec}$ for
$Be^7$ neutrino and in a charged current event, which occurs
incoherently,  a rate is about one hundred per year in Super Kamiokande.  A
mean free path of 1 MeV neutrino in the earth is about $4 \times
10^{10}$ Km and is slightly smaller in the moon. The earth and the moon are 
quite transparent as far as the charged current interaction is concerned. 
Strength of the neutral current interaction is smaller than the charged
current interaction and  elastic scattering process 
$\nu_e+e^{-} \rightarrow e^{-}+\nu_e$ is similar to the neutral current
interaction but has a larger strength. In these processes  the neutrino
and other matters can stay in almost the same quantum states, and  by
the coherent sum of amplitudes a total amplitude is enhanced. This
effect becomes important when neutrino interacts with a massive object
like a star. The 
moon is a star around the earth and can be located between the earth and
the sun.  So the coherent interaction of the solar neutrino with the
moon can take place. We study  the coherent scattering of the solar
neutrino with moon in this paper.

The MSW coherent matter effect Ref.[2-3] is actually one of the
important ingredients in understanding current neutrino oscillation 
experiments. Because expected day-night effect has not been observed,
the MSW effect would not be important for the earth and for the moon.     
Double slitt like effect we study in this paper is a spatial
interference effect of the lightest neutrino. This effect is sensitive
to the absolute magnitude of the lightest neutrino mass.

The solar neutrino has an energy  in MeV region 
and the coresponding de Broglie wave length is $10^{-13}m $ ,whearas the 
radius of the moon is of order $10^6$ m. Because the wave length is much 
smaller than the total size, this regime may be thought as a geometrical 
optics regime or the classical regime for the neutrino where the wave 
character is normally washed out. However the interaction strength of 
neutrino with matter is extremely weak and the neutrino propagates
almost freely even in matter with keeping an original phase. 
Consequently the propagation of neutrino in the moon should be treated
by quantum mechanical wave equation even in the macroscopic regime. 
Interference is a key phenomenon of wave dynamics. We show that to
generate interference for solar neutrino, wave packet character
,which have been studied for neutrino oscillations in Ref. [4-8], is
important. Using reasonable parameter values, we estimate the magnitude
of the effect from overlapp between an initial wave packet and a final
wave packet. We show that  an observation of the effect would be possible
if the absolute value of the neutrino mass is $10^{-4} eV$ or less.
Observation of  the interference should verify quantum mechanics in Gm 
region and give new informations on the neutrino mass
and on the interior of the moon. \\

------------------------------------------------------\\
ishikawa@particle.sci.hokudai.ac.jp(K. Ishikawa)  
\newpage
By replacing  the quark pair operators or the electron pair operators in
the four Fermi interaction with the density of electrons and the
quarks, we obtain  a potential term in Dirac equation. We solve the
equation and find that  the effect of the potential arises  in the
momentum dependent phase factor. Although the effect disappers 
from the magnitude of a plane wave, the potential gives an interference
effect to the wave packet since  the wave packet 
is a linear combination of plane waves. 
Interference among the plane waves is  generated and an observable 
effect is obtained. It is predicted that the solar neutrino flux 
changes with time when they are blocked by the moon during eclipses.    

{\bf potential }

Neutrino interacts with quarks $q(x)$ or with the electron $e(x)$ in matter
by the charged current interaction
\begin{equation}
{G_{F} \over \sqrt 2} \bar\nu(x)\gamma_{\mu}(1 -\gamma_5) e(x) \{\bar d(x) \gamma^{\mu}(1 -\gamma_5) u(x) +\bar e(x) \gamma^{\mu}(1 -\gamma_5) \nu(x) \},
\label{eq:chargei}
\end{equation}

and by the neutral current interaction,
\begin{eqnarray}
& &{G_{F} \over \sqrt 2} \bar\nu(x)(1 -\gamma_5)\gamma_{\mu} \nu(x) \{ g_V(u)
\bar u(x) \gamma^{\mu}u(x) + g_V(d)\bar d(x) \gamma^{\mu}d(x)  \\
& &+ g_V(e) \bar e(x) \gamma^{\mu}e(x) + axial vectors \}. \nonumber \\
\nonumber
\label{eq:neutrali}
\end{eqnarray}
Throughout this paper we  assume that the electron neutrino is lightest
and the mixing angle is negligible.
The      second term in charged current interction 
has a same final state and initial state as electron neutrino
neutral current interaction 
and by Fiertz transformation the interaction is written as
\begin{equation}
{G_{F} \over \sqrt 2} \bar\nu(x) (1 -\gamma_5)\gamma_{\mu} \nu(x)\{ 
\bar e(x) \gamma^{\mu}e(x) +axial vector \}.
\end{equation}
The coefficents $g_V(u),g_V(d),g_V(e)$ are given by the Weinberg angle
$\theta_W$ as,
\begin{eqnarray}
& &g_V(u)= {1 \over 2}-2 \times{2 \over 3}\times{\sin}^2{\theta_W}   \\    
& &g_V(d)=  -{1 \over 2}-2 \times(-{1 \over 3})\times{\sin}^2{\theta_W} \\  
& &g_V(e)=  -{1 \over 2}+2 \times{\sin}^2{\theta_W}
\label{eq:neu}   
\end{eqnarray}  
Using the current value $\sin^2 {\theta_W}=0.23108 $ [9], we have
\begin{equation}
 g_V(u)=0.19, g_V(d)= -0.35,g_V(e)=-0.034
\end{equation}

The charged current interaction,Eq.(\ref{eq:chargei}), transforms the neutrino to
the electron. Since the final state in the matter is different from the 
initial state, the process occurs incoherently. The mean free path of the
neutrino in the earth from the incoherent scattering $\nu+n \rightarrow
e+p$
is computed from 
the crossection, $\sigma$, and the neutron density $\rho_n$. 
By  using the  crosssection and  the density,      
\begin{eqnarray}
& &\sigma=   {{G_F}^2 \over \pi} 2M_n E_{\nu}   \\
& & \rho_n= 0.556{r \over 2} N_{Avo}/{cm}^3,
\end{eqnarray}
where $r$ is the specific gravity and $N_{Avo}$ is Avogadoro Number, we
have  the mean free path, 
\begin{eqnarray}
l&=&{1 \over \sigma \rho} \nonumber \\
 &=&10^{11} Km 
\end{eqnarray}
for $r=5$ and $E_{\nu}=1$ Mev. Hence the mean free path is much larger
than the size of the earth, $10^5 Km $ or the size of the moon, $2\times 10^3$ Km.  Neutrino propages in the earth or in the moon almost
freely as far as the charged current interactions are concerned.
 
%\subsection{Coherent scattering}
The strengths of neutral current interactions, Eq.(\ref{eq:neutrali}),
are about the same
as the charged current interactions. So a probability of incoherent 
scattering event from the neutral current interaction is also about the same
as the charged current interaction and is very small. 

Coherent scattering is possible in the neutral current interactions
since the final state in the matter is the same as the initial state. In
the coherent scattering, amplitude from each atom is coherently added
and a total amplitude can become much larger than the original
amplitude. In this case, an enhancement is possible. We will see
that the charge of vector current can have a coherent contribution.

To study weak matrix elements, expectation value of the current in the
matter
\begin{eqnarray}
& &\langle Matter|\bar f(0)\gamma_{\mu}\Gamma f(0)|Matter\rangle\\
& &=\sum_i\bar f_i\gamma_{\mu}\Gamma f_i(0)\nonumber
\end{eqnarray}
is needed where summation over atoms is taken. Expectation values of the 
currents in one particle state with a momentum ${\vec p}$ and a spin 
${\vec s}$ are 
\begin{eqnarray}
& &\bar u(p,s) \gamma_0 u(p,s)={p_0 \over m},\\
& & \bar u(p,s) \gamma_i u(p,s)={p_i \over m},\nonumber \\
& &\bar u(p,s) \gamma_0 \gamma _5 u(p,s)= 0, \\
& & \bar u(p,s) \gamma_i \gamma_5 u(p,s)=s_i.\nonumber 
\end{eqnarray} 
Only the zeroth component of the vector charge is positive definite and 
nearly 1 in the forward scattering region of the neutrinos.
Other components are not positive definite and are small. Hence after the summation over whole atoms, 
only the vector charge becomes non-zero value $ \sum _i \bar u(p,s)
\gamma_0 u(p,s)=\rho $ and others vanish.

Consequently coherent neutrino interaction with matter is obtained by
the vector charge and
is reduced to  an effective two body Hamiltonian,
\begin{eqnarray}
& &\bar\nu(x){1-\gamma_5 \over 2}\gamma_{\mu} \nu(x) V_0(x),\\
& &V_0(x)= \sqrt 2 G_F (\rho_e(x) -0.5 \rho_{neutron}(x)), 
\label{eq:potential}
\end{eqnarray}
where $\rho_e(x)$ is the density of the electrons and $\rho_{neutron}(x)$ is
the density of the neutron .

{\bf plane wave}

We solve a Dirac equation with the potential term that is produced by  the 
interaction with matter. Since the potential is caused by the weak
interaction, the magnitude is very small and is proportional to the
Fermi coupling constant. The range of the potential we study,on the
other hand,  is very large. So it is interesting to see if the effect of 
the potential is observable.  

Let us solve a Dirac equation with a left-handed potential term, 
\begin{equation}
i\hbar {\partial \over \partial t} \psi(x)=(i {\vec \alpha}\cdot{\vec p} +m\beta) \psi(x) +V(x)({1-\gamma_5 \over 2}) \psi(x), 
\end{equation}
\begin{eqnarray}
& &V(x)=V_0,r \le R \\
& & V(x)= 0,r \ge R, \nonumber
\end{eqnarray}
here $R$ is a large macroscopic value and $V_0$ is a small value.
We will see that the product $V_0 R$ is order 1.  
 
We obtain a stationary solution of the energy $E$ 
\begin{equation}
\psi(x)=\exp{( E t / i\hbar)}\psi({\vec x})
\end{equation}
with a boundary condition at $z \rightarrow -\infty$
\begin{equation}
\phi({\vec x})=e^{i {\vec k}\cdot {\vec x} }u({\vec k}),
\end{equation}
where $u({\vec k})$ is a free Dirac spinor of a momentum ${\vec k}$.
$\psi({\vec x})$ satisfies an intergral equation,
\begin{equation}
\psi({\vec x})=\phi({\vec x})+\int d{\vec x}'D({\vec x}-{\vec x}')  V({\vec x}')({1-\gamma_5 \over 2}) \psi({\vec x}'),
\end{equation}
with the Green function of the Dirac operator, 
\begin{equation}
D({\vec x}-{\vec x}') 
=\int {d{\vec p} \over (2\pi)^3} e^{i {\vec p} \cdot ({\vec x}-{\vec x}')} {{E+{\vec \alpha} \cdot {\vec p}+m\beta }  \over {E^2-  {\vec p}^2 -m^2+i\epsilon }}. 
\end{equation}
By applying Born approximation,we have an solution 
\begin{eqnarray}
\psi({\vec x})&=&\phi({\vec x})+\psi({\vec x})^{(1)}+\psi({\vec x})^{(2)}+
\psi({\vec x})^{(3)}+-, \\
\psi({\vec x})^{(l+1)}&=&\int d{\vec x}'D({\vec x}-{\vec x}')  V({\vec x}')({1-\gamma_5 \over 2}) \psi({\vec x}')^{(l)},\nonumber  \\
\psi({\vec x})^{(1)} &=&e^{i{\vec k}\cdot {\vec x}}\int {d{\vec p} \over (2\pi)^3}e^{i({\vec p}-{\vec k})\cdot {\vec x}}{1 \over {\vec k}^2-{\vec p}^2+i\epsilon }\times \\
& &({1-\gamma_5 \over 2}( 2E({\vec k})+{\vec \alpha}\cdot({\vec p}-{\vec k}))+\gamma_5\beta
m)u({\vec k}) \tilde V({\vec k}-{\vec p})\nonumber 
\label{eq:persol}
, 
\end{eqnarray}
where 
\begin{eqnarray}
\tilde V({\vec k}-{\vec p})&=&\int d{\vec x}'e^{i({\vec k}-{\vec p})\cdot{\vec x}'}V({\vec x}')  \\
 &=&4\pi V_0 R^3 v_0(q), \nonumber \\
v_0(q)&=&{1 \over q}({\sin q \over q^2}-{\cos q \over q} ),\\
q&=&R|({\vec p}-{\vec k})|  \nonumber
\end{eqnarray}
Identity,
\begin{eqnarray}
& &(E({\vec k})+{\vec \alpha} \cdot {\vec p}+m\beta){1-\gamma_5 \over 2} u({\vec k}) \nonumber \\
& &={1-\gamma_5 \over 2} 2E({\vec k})u({\vec k})+\gamma_5m\beta u({\vec k})
+{1-\gamma_5 \over 2} {\vec \alpha}\cdot{({\vec p}-{\vec k})} u({\vec k})
\end{eqnarray}
was used in Eq.(\ref{eq:persol}). Finally we have 
\begin{eqnarray}
\label{eq:solution}
& &\psi({\vec x})^{(1)}=e^{i{\vec k}\cdot {\vec x}}\{ {1-\gamma_5 \over 2}2E({\vec k}) {V_0 R \over 2\pi^2}u({\vec k})F({\vec x},{\vec k}) \\
& &+\gamma_5\beta m {V_0 R \over 2\pi^2}u({\vec k})F({\vec x},{\vec k})+{1-\gamma_5 \over 2 }{V_0 \over 2\pi^2}{ \alpha}_i F_i({\vec x},{\vec k})u({\vec k})\},
\nonumber \\
\label{eq:function1}
& &F({\vec x},{\vec k})=\int d{\vec q}{1 \over -2{\vec k}\cdot {\vec q}+i\epsilon}v_0(q)e^{i{\vec q}\cdot{\vec x} \over R}\\
\label{eq:function2}
& &F_i({\vec x},{\vec k})=\int d{\vec q}q_i{1 \over -2{\vec k}\cdot {\vec q+i\epsilon}}
v_0( q)e^{i{\vec q}\cdot{\vec x} \over R}.
\end{eqnarray}
Eqs. (\ref{eq:function1}) and (\ref{eq:function2}) become 
\begin{eqnarray}
& &F({\vec x},{\vec k})= i{4 {\pi}^2 \over  2k}\sqrt{1-{\vec \xi}_T^2},\\
& &F_i({\vec x},{\vec k})={\partial \over \partial k_i} F({\vec x},{\vec k})
%-i{4 {\pi}^2 \over  2k} {\xi_{i} \over \sqrt{1-{\xi}^2}}
\end{eqnarray}
in the region ${\vec k}\cdot{\vec x} > R,1>{\vec \xi_T}^2$ where 
${\vec \xi}={\vec x}/R,{\vec \xi}_{T}={\vec \xi}-{\vec k}({\vec
k}\cdot{\vec \xi})/{k^2}$. In the region where $R$ and ${E \over m} $
are large, the first term is dominant over the second
term and the third term in Eq.(\ref{eq:solution}),
and we have
\begin{equation}
\psi({\vec x})^{(1)}=   {1-\gamma_5 \over 2}2E({\vec k}) {V_0 R \over 2\pi^2}F({\vec x},{\vec k}) \phi({\vec x}). 
\end{equation} 
Thus the first order term changes with position in the range R even though
the wave length is microscopic size. It should be noted however that
this term is pure imaginary.  It is suggested  that 
correction terms modify only phase factor of the wave function. We show in the
following that this is the case in fact.
    
 The second order term is  computed in a similar manner. The dominant
 term is given as,
\begin{equation}
\psi({\vec x})^{(2)}= 
({1-\gamma_5 \over 2}2E({\vec k}) {V_0 R \over 2\pi^2})^2 F^{(2)}({\vec x},{\vec k})  \phi({\vec x}), 
\end{equation} 
where the coefficient is given as 
\begin{equation}
F^{(2)}({\vec x},{\vec k})=\int d{\vec q}_{1}d{\vec q}_{2} {1 \over -2{\vec k}\cdot ({\vec q}_{1}+{\vec q}_{2})+i\epsilon}v_0(q_1) {1 \over -2{\vec k}\cdot {\vec q}_{2}+i\epsilon}v_0(q_2)e^{i( {\vec q}_{1}+ {\vec q}_{2})\cdot{\vec x} \over R}.
\end{equation}
By  writing the integral with a symmetric manner, we have
\begin{eqnarray}
& &F^{(2)}({\vec x},{\vec k}) \\
& &={1 \over 2!}\int d{\vec q}_{1}d{\vec q}_{2} {1 \over -2{\vec k}\cdot ({\vec q}_{1}+{\vec q}_{2})+i\epsilon}
({1 \over -2{\vec k}\cdot {\vec q}_{2}+i\epsilon}+ {1 \over -2{\vec k}\cdot {\vec q}_{1}+i\epsilon})\nonumber\\
& &\times v_0(q_1) v_0(q_2)e^{i( {\vec q}_{1}+ {\vec q}_{2})\cdot{\vec x} \over R}\nonumber \\
& &={1 \over 2!}\int d{\vec q}_{1}d{\vec q}_{2} {1 \over -2{\vec k}\cdot{\vec q}_{1}+i\epsilon} 
 {1 \over -2{\vec k}\cdot {\vec q}_{2}+i\epsilon} v_0(q_1)
v_0(q_2)e^{i( {\vec q}_{1}+ {\vec q}_{2})\cdot{\vec x} \over R}\nonumber \\
& &={1 \over 2!}{F({\vec x},{\vec k})}^2\nonumber.
\end{eqnarray}
Thus the second order term becomes,
\begin{equation}
\psi({\vec x})^{(2)}={1 \over 2!} ({1-\gamma_5 \over 2}2E({\vec k}) {V_0 R \over 2\pi^2}F({\vec x},{\vec k}))^2 \phi({\vec x}) 
\end{equation}
and higher order terms are written in the same manner
\begin{equation}
\psi({\vec x})^{(l)}={1 \over l!} ({1-\gamma_5 \over 2}2E({\vec k}) {V_0 R \over 2\pi^2}F({\vec x},{\vec k}))^l \phi({\vec x}). 
\end{equation}
Adding all higher order terms we have the wave function 
\begin{equation}
\label{eq:fsolution}
\psi({\vec x})= \exp{(i{\vec k}\cdot{\vec x} + i {1-\gamma_5 \over 2}2 V_0{1 \over k} \sqrt { {\vec k}^2(R^2-{\vec x}^2)+{({\vec k}\cdot{\vec x})^2}})}u({\vec k}) .
\end{equation} 

 Substituing the Fermi coupling constant, the radius and the
 density of the moon assuming the specific gravity $r$,
\begin{eqnarray}
& &G_F=1.16 \times 10^{-5} (GeV)^{-2}       \\
& & R= 1.74 \times 10^3 Km          \\
& &\rho_e=\rho_{neutron}=0.556 {r \over 2} N_{Avo}/{cm}^3
\end{eqnarray}
to Eq.(\ref{eq:potential}), the numerical constant in
Eq.(\ref{eq:fsolution}) becomes for  $r=5$ 
\begin{equation}
2 V_0 R= 0.95.
\end{equation}
This is an interesting value to see an interference effect. However the 
correction is in the phase factor and disappears in the  $|\psi({\vec
x})|^2$. So the neutrino flux is the same as free wave if the neutrino
is a plane wave. It is impossible to observe the interference effect
using any plane wave in the present situation.  

{ \bf wave packet}

Realistic solar neutrino is not a plane wave but is a linear combination 
of plane waves, a finite wave packet Ref.[4-9,11-12]. The scattering amplitude
is the overlapp between the initial wave packet and the final wave
packet. The former is determined from the beam and the latter is
determined from the detector.  

From the Eq.(\ref{eq:fsolution}), 
the phase of the initial wave function through the moon depends on several
parameters such as momentum and position and the wave packet is certain
linear combinations of these waves. We will show that the overlapp
between these wave packets and the final wave packets have interference 
effect if the mass of the lightest neutrino is about $10^{-4} eV/c^2 $ or less.

A wave packet expands during a propagation [12,13]. Speed of expansion is
determined by  the velocity variance and is  dominant in the transverse 
direction for a relativistic particle. Using the initial size $\delta
x$, the maxmimum velocity in the transverse direction is given as 
$ v_T={\delta P_T \over E} = {\hbar \over {\delta x} E}$. 
The size in the transverse direction is a product of the velocity
and the propagation time  
\begin{equation}
\Delta x_T=v_T \delta t.
\end{equation} 
The size for the neutrino of $E=0.6MeV,{\delta x}=10^{-10}m $  becomes  
$3\times10^5 Km$ for $\delta t=500s $ and $10^3 Km$ for $\delta
t=1s$. The former one is  the packet size when it propagates
from the sun to the earth and the latter is the packet size when it 
propagates from the earth to the moon. The radius of moon is 
$1.7 \times 10^3 Km$ and is about the same as the packet size of the
neutrino at the moon which is observed at the earth with a microscopic size. 
So the wave packet effects are relavant for the solar neutrino that is
detected at the earth.

 The Gaussian wave packet of the variance $\sigma \hbar$ at $t=0,{\vec x}={\vec X_0}$ expands at a much 
 later time $t>>t_0$ or at a previous time $t<<-t_0$ and behaves at a
 distant position ${\vec x}$ as 
\begin{eqnarray}
& &\psi({\vec x},t)= N e^{-i{ {mc \over \hbar}\sqrt {(ct)^2-({\vec x}-{\vec X_0})^2} } -{1 \over 2 \sigma \hbar}({\vec P}_X-{\vec p}_0)^2},\\ 
& &{\vec P}_X=mc{1 \over \sqrt{(ct)^2-({\vec x}-{\vec X}_0)^2}} ({\vec x}-{\vec X}_0), \nonumber
\end{eqnarray}    
where N is a normalization factor and  ${\vec p}_0$ is the
central value of the momentum. From the Gaussian term
the wave packet size is determined. The phase factor is written by the
use of the momentum as,
\begin{equation}
\phi={ (mc^2)^2 \over \hbar |{\vec P}_{X} c |} t.
\end{equation}
The phase factor $F({\vec k},{\vec x})$ is added for the neutrino which 
penetrates inside the moon.  

We study the overlapp of two wave packets at a certain time at around
the moon. Initial wave packet which is emitted at a time $T_1$ [14] in
the sun and propagates toward the earth through the Moon and final wave packet
which is detected at a time $T_2$ at the detector  are located around the
 Moon at a time $t=T_2-\delta t $ where $\delta t$ is around one
 second. The function $F({\vec k},{\vec x})$ is in
 one of the wave function and becomes smooth function of ${\vec x}$ in this
region and computation is straightforward.

The initial wave packet which is emitted from  the sun at $({\vec
X}_1,T_1)$ behaves at  $({\vec x},t)$ as,
\begin{eqnarray}
\label{eq:inwave}
& &\psi_{in}({\vec x},t)= N e^{-i{ {mc \over \hbar}\sqrt {(c(t-T_1))^2-({\vec x}-{\vec X_1})^2} } -{1 \over 2 \sigma \hbar}({\vec P}_{X_1}-{\vec p^{in}}_0)^2},\\ 
& &{\vec P}_{X_1}=mc{1 \over \sqrt{(c(t-T_1))^2-({\vec x}-{\vec X}_1)^2}} ({\vec x}-{\vec X}_1). \nonumber
\end{eqnarray}    
The final wave packet detected at $({\vec
X_2},T_2)$ behaves at the same $({\vec x},t)$ as,
\begin{eqnarray}
\label{eq:outwave}
& &\psi_{out}({\vec x},t)= N e^{-i{ {mc \over \hbar}\sqrt {(c(T_2-t))^2-({\vec x}-{\vec X_2})^2} } -{1 \over 2 \sigma \hbar}({\vec P}_{X_2}-{\vec p}_0^{out})^2+F({\vec x})},\\ 
& &{\vec P}_{X_2}=mc{1 \over \sqrt{(c(T_2-t))^2-({\vec x}-{\vec X}_2)^2}} ({\vec x}-{\vec X}_2). \nonumber,
\end{eqnarray}   
We have chosen the situation where the wave front of the initial wave
packet does not reach the Moon but the wave front of the final wave
packet reaches the Moon. The phase in Eqs.(\ref{eq:inwave}) and 
(\ref{eq:outwave} ) depend on the absolute value of the neutrino
mass and momentum.  For a neutrino of the mass $10^{-4}eV/c^2$ and the momentum $1MeV/c$,
the phase factor $\phi$ is written as 
\begin{eqnarray}
& &\phi=5 {|{\vec x}| \over x_0} \times 10^{-3}, \\
& & x_0=1000Km, \nonumber
\end{eqnarray}
and becomes negligbly small in the scale of Moon.
So the phase factor of ${\psi_{in}({\vec x})}$ is constant $\phi_0$ and
the phase factor of ${\psi_{out}({\vec x})}$ is negligibly small in the same
scale. Gaussian factors are almost one in the same region.
Then overlapp of the two wave functions is given as,
\begin{equation}
(\psi_{out}({\vec x},t), \psi_{in}({\vec x},t))=e^{i\phi_0} \int d {\vec r} N e^{iF( {\vec r}) -{1 \over 2 \sigma \hbar}({\vec P}_{X_1}-{\vec p^{in}}_0)^2-{1 \over 2 \sigma \hbar}({\vec P}_{X_2}-{\vec p^{out}}_0)^2}.
\end{equation}  
Since the Gaussian terms are positive definite the integral is reduced
by the effect of phase factor $F({\vec x})$. The reduction rate depends
on parameters such as wave packet sizes, the energy and others. Using
the Moon radius for the size of detecting wave packet around Moon we
have the neutrino flux at about $0.85 $  of the normal  flux value  
for ${\vec p}_0^{in}={\vec p}_0^{out}={\vec P}_X $ at a total solar eclipse. 
Reduction is almost the same for other momenta.

%\section{Summary}
We  studied coherent scattering of solar neutrino by the moon and found
that  the moon modifies the phase of solar neutrino wave function. The
result is surprizing from two reasons. First one is extreme weakness 
of potential and second one is the fact that the wave character is seen 
in extremely large scale.  
They  became possible  because, despite extreme weakness of  potential, the
volume of finite potential region is very large and the product of two 
values is of order one. The other feature is seen from the fact that the 
ratio between  the distance and de Broglie wave length is 
around $10^{20}$. Normally in this regime the wave character are washed
away due to a rapid oscillation of phase and geometrical optics regime
is realized. Interference should have not been produced in this
case. However in the present case, interference is in fact produced. This
is because  the time dependent phase and space dependent phase of the
relativistic waves cancell and  consequently the phase difference in
large distance is not washed away and wave phenomena occurs. 

As an implication, we suggests that the time dependent moduration
of solar neutrino flux occur during eclipse if the suitable detector is
used. If a macroscopic wave phenomenon of neutrino is verified, this  can
be used for testing  quantum mechanics in the scale of Gmeter range. 
New informations on the absolute  value of the lightest neutrino mass and the 
interior of the moon could be supplied also. 
%%%%%%%%%%%%%%%%%%%%%%%%%%%%%%%%%%%%%%%%%%%%%%%%%%%%%%%%%%%%%%%%%%%%%%%%%%%

{\bf Acknowledgements}

This work was partially supported by the special Grant-in-Aid
for Promotion of Education and Science in Hokkaido University
provided by the Ministry of Education, Science, Sports and Culture,
 a Grant-in-Aid for Scientfic Research on Priority Area ( Dynamics of
Superstrins and Field Theories, Grant No. 13135201), and a Grant-in-Aid 
for Scientfic Research on Priority Area ( Progress in Elementary
Particle Physics of the 21st Century through Discoveries  of Higgs Boson and
Supersymmetry, Grant No. 16081201) provided by 
the Ministry of Education, Science, Sports and Culture, Japan.
K. I. thanks G. Takeda for informing Ref. [15].

\textbf{References}

\bigskip

[1] T. Hirata et al, Phy. Rev. Lett. \textbf{58},
(1987)1490; Phys.Rev.\textbf{D38}, (1988)448.

[2] L. Wolfenstein, Phys. Rev. \textbf{D17}, (1978)2369

[3] S. P. Mikheyev,and A. Yu. Smirnov, Yad. Fiz.\textbf{42}, (1985)1441

[4] B. Kayser, Phys. Rev. \textbf{D24}, (1981)110;
Nucl. Phys. \textbf{B19}(Proc.Suppl), (1991)177

[5] C. Giunti, C. W. Kim, and U. W. Lee, Phys. Rev. \textbf{D44}, (1991)3635

[6] S. Nussinov, Phys. Lett. \textbf{B63}, (1976)201

[7] K. Kiers ,N. Nussinov and N. Weisis, Phys. Rev. \textbf{D53}, (1996)537.

[8] C. Y. Cardall, Phys. Rev.  \textbf{D61}, (2000)073006

[9] M. Beuthe, Phys. Rev.  \textbf{D66},  (2002)013003;

[10] Particle Data table, S. Eidelman et al. , Physics Letters
\textbf{B592}, (2004)1

[11] A. Asahara, K.Ishikawa, T. Shimomura, and T. Yabuki,
Prog. Theor. Phys. \textbf{113}, (2005)385; T. Yabuki and K. Ishikawa,
Prog. Theor. Phys. \textbf{108}, (2002)347.

[12] K. Ishikawa and T. Shimomura ,''Generalized S-matrix in Mixed 
Representation ''Hokkaido University preprint  hep-ph/0508303

[13] M. L. Goldberger and Kenneth M. Watson, ``Collision Theory ``
(John Wiley \& Sons, Inc. New York, 1965).

[14] Neutrino production time is neither known nor defined with a
definite value because K-electron capture rate of $Be^7$ is so slow.
Mean free time of $Be^7$ core or $Li^7$ core may be large. Also no
observation is made on these objects  and  time width becomes
large. See [12] on this point. Consequently we simply take a linear
combination of the different time. 

[15]Coherent interactions of neutrinos with matters was studied in
different context by, P. Langacker, J. P. Leveille, and J. Sheinen,
 Phys. Rev.  \textbf{D27},  (1983)1228. 

[16] After we have completed our paper we found the following
paper:Solar Neutrinos and the Eclipse Effect, M. Narayan, G. Rajasekaran,
and R. Sinha,and C. P. Burgess,  Phys. Rev.  \textbf{D60}, (1999)073006,
which discusses the eclipse effect in MSW solar neutrino oscillations.

\end{document}